\documentclass[10pt]{article}
\usepackage{epsfig,graphicx}
\usepackage{fpcp04}

\def\etal         {\em et al.}
\def\to           {\ensuremath{\rightarrow}}
\def\B            {\ensuremath{B}}
\def\bb           {\ensuremath{B\overline{B}}}
\def\Bz           {\ensuremath{B^{0}}}
\def\Bzb          {\ensuremath{\overline{B}^{0}}}

\def\Br           {\ensuremath{{\cal B}}}

\def\CP      {\ensuremath{CP}}
\def\T       {\ensuremath{T}}
\def\CPV     {\ensuremath{CPV}}

\def\C       {\ensuremath{C}}

\def\acp     {\ensuremath{{\cal A}_{CP}}}

\def\fL      {\ensuremath{f_L}}
\def\ptrue   {\fL}
\def\mes     {M_{ES}}

\def\deltae  {\ensuremath{\Delta E}}

\def\epem      {\ensuremath{{e^+e^-}}}
\def\qqbar     {\ensuremath{{q \overline{q}}}}

\newcommand{\e}      [1]  {\ensuremath{\times 10^{#1}}}
\newcommand{\su}     [1]  {\ensuremath{SU(#1)}}
\def\aone        {\ensuremath{a_1}}
\def\Bztoaonepi {\ensuremath{\Bz\to \aone^\pm \pi^\mp}}
\def\belle     {Belle}

\def\beq{\begin{equation}}
\def\eeq#1{\label{#1}\end{equation}}
\def\eeqn{\end{equation}}

\def\beqa{\begin{eqnarray}}
\def\eeqa#1{\label{#1}\end{eqnarray}}
\def\eeqan{\end{eqnarray}}

\let\bar=\overbar

\def\etal{{\it et al.}}

\def\Dslash{\not{\hbox{\kern-4pt $D$}}}
\def\dslash{\not{\hbox{\kern-2pt $\del$}}}

\def\msb{{\bar{\ssstyle M \kern -1pt S}}}

\def\BB0bar{B^0 {\overline B}^0}
\def\BB0dbar{B_d^0 {\overline B}_d^0}
\def\BB0sbar{B_s^0 {\overline B}_s^0}

\RequirePackage{xspace}

\usepackage{relsize}
\def\babar{\mbox{\slshape B\kern-0.1em{\smaller A}\kern-0.1em
    B\kern-0.1em{\smaller A\kern-0.2em R}}}

\def\epem       {\ensuremath{e^+e^-}\xspace}

\def\qqbar {\ensuremath{q\overline q}\xspace}

\def\Kbar  {\kern 0.2em\overline{\kern -0.2em K}{}\xspace}

\def\Kz    {\ensuremath{K^0}\xspace}
\def\Kzb   {\ensuremath{\Kbar^0}\xspace}
\def\KzKzb {\ensuremath{\Kz \kern -0.16em \Kzb}\xspace}
\def\Kp    {\ensuremath{K^+}\xspace}
\def\Km    {\ensuremath{K^-}\xspace}

\def\KpKm  {\ensuremath{\Kp \kern -0.16em \Km}\xspace}

\def\Dbar    {\kern 0.2em\overline{\kern -0.2em D}{}\xspace}

\def\Dz      {\ensuremath{D^0}\xspace}
\def\Dzb     {\ensuremath{\Dbar^0}\xspace}
\def\DzDzb   {\ensuremath{\Dz {\kern -0.16em \Dzb}}\xspace}
\def\Dp      {\ensuremath{D^+}\xspace}
\def\Dm      {\ensuremath{D^-}\xspace}

\def\DpDm    {\ensuremath{\Dp {\kern -0.16em \Dm}}\xspace}

\def\B       {\ensuremath{B}\xspace}
\def\Bbar    {\kern 0.18em\overline{\kern -0.18em B}{}\xspace}

\def\BB      {\ensuremath{B\Bbar}\xspace} 
\def\Bz      {\ensuremath{B^0}\xspace}
\def\Bzb     {\ensuremath{\Bbar^0}\xspace}
\def\BzBzb   {\ensuremath{\Bz {\kern -0.16em \Bzb}}\xspace}
\def\Bu      {\ensuremath{B^+}\xspace}
\def\Bub     {\ensuremath{B^-}\xspace}

\def\BpBm    {\ensuremath{\Bu {\kern -0.16em \Bub}}\xspace}

\mathchardef\Upsilon="7107
\def\Y#1S{\ensuremath{\Upsilon{(#1S)}}\xspace}% no space before {...}!

\mathchardef\Deltares="7101
\mathchardef\Xi="7104
\mathchardef\Lambda="7103
\mathchardef\Sigma="7106
\mathchardef\Omega="710A

\def\Deltabar{\kern 0.25em\overline{\kern -0.25em \Deltares}{}\xspace}
\def\Lbar{\kern 0.2em\overline{\kern -0.2em\Lambda\kern 0.05em}\kern-0.05em{}\xspace}
\def\Sigbar{\kern 0.2em\overline{\kern -0.2em \Sigma}{}\xspace}
\def\Xibar{\kern 0.2em\overline{\kern -0.2em \Xi}{}\xspace}
\def\Obar{\kern 0.2em\overline{\kern -0.2em \Omega}{}\xspace}
\def\Nbar{\kern 0.2em\overline{\kern -0.2em N}{}\xspace}
\def\Xb{\kern 0.2em\overline{\kern -0.2em X}{}\xspace}

\def\mes        {\mbox{$m_{\rm ES}$}\xspace}

\newcommand{\tev}{\ensuremath{\mathrm{\,Te\kern -0.1em V}}\xspace}
\newcommand{\gev}{\ensuremath{\mathrm{\,Ge\kern -0.1em V}}\xspace}
\newcommand{\mev}{\ensuremath{\mathrm{\,Me\kern -0.1em V}}\xspace}
\newcommand{\kev}{\ensuremath{\mathrm{\,ke\kern -0.1em V}}\xspace}
\newcommand{\ev}{\ensuremath{\mathrm{\,e\kern -0.1em V}}\xspace}
\newcommand{\gevc}{\ensuremath{{\mathrm{\,Ge\kern -0.1em V\!/}c}}\xspace}
\newcommand{\mevc}{\ensuremath{{\mathrm{\,Me\kern -0.1em V\!/}c}}\xspace}
\newcommand{\gevcc}{\ensuremath{{\mathrm{\,Ge\kern -0.1em V\!/}c^2}}\xspace}
\newcommand{\mevcc}{\ensuremath{{\mathrm{\,Me\kern -0.1em V\!/}c^2}}\xspace}
\def\mus  {\ensuremath{\rm \,\mus}\xspace}

\def\mus        {\ensuremath{\,\mu{\rm s}}\xspace}    %% microsecond
\def\to                 {\ensuremath{\rightarrow}\xspace}

\newcommand{\stat}{\ensuremath{\mathrm{(stat)}}\xspace}
\newcommand{\syst}{\ensuremath{\mathrm{(syst)}}\xspace}

\def\pep2{PEP-II}

\def\gsim{{~\raise.15em\hbox{$>$}\kern-.85em
          \lower.35em\hbox{$\sim$}~}\xspace}
\def\lsim{{~\raise.15em\hbox{$<$}\kern-.85em
          \lower.35em\hbox{$\sim$}~}\xspace}

\def\CP                {\ensuremath{C\!P}\xspace}
\xspace

\newcommand{\jprlBase}       {Phys.\ Rev.\ Lett.\xspace}
\newcommand{\jprBase}        {Phys.\ Rev.\xspace}
\newcommand{\jplBase}        {Phys.\ Lett.\xspace}

\newcommand{\nimBaseC}       {Nucl.\ Instr.\ and Methods\xspace}

\newcommand{\nima}      [1]  {\nimBaseC~A~{\bf #1}}

\newcommand{\plb}       [1]  {\jplBase\ B~{\bf #1}}

\newcommand{\jprl}      [1]  {\jprlBase\ {\bf #1}}
\newcommand{\jprd}      [1]  {\jprBase\ D~{\bf #1}}

\newcommand{\progtp}    [1]  {{Prog.\ Th.\ Phys.\ {\bf #1}}}

\def\jetset74   {\mbox{\tt Jetset \hspace{-0.5em}7.\hspace{-0.2em}4}\xspace}
\begin{document}
\begin{flushleft}
SLAC-PUB-11888
\end{flushleft}

\Title{Rare hadronic \B\ decays.}
\bigskip

%%%%%%%%%%%%%%%%%%%%%%%%%%%%%%%%%%%%%
% Label to flag the first page of your contribution
% Replace Perret by your name starting with a capital letter
%
\label{AJBevanStart}

%%%%%%%%%%%%%%%%%%%%%%%%%%%%%%%%%%%%%
% Your name
%
\author{Adrian Bevan\index{Bevan, A. J.} }
%%%%%%%%%%%%%%%%%%%%%%%%%%%%%%%%%%%%%
% Your address
%
\address{Department of Physics\\
Queen Mary, University of London, Mile End Road, E1 2NS, UK. \\e-mail: {\em bevan@slac.stanford.edu}\\
%Liverpool University, Liverpool, UK.\\
%(on the behalf of the \babar\ and Belle Collaborations.)\\
Presented at the International Conference on Flavor Physics,\\ Chung-Li, Taiwan, $3^{rd}-8^{th}$ October 2005.\\
}

\makeauthor\abstracts { Rare hadronic \B-meson decays allow us to
study \CP\ violation. The class of \B-decays final states
containing two vector mesons provides a rich set of angular
correlation observables to study. This article reviews some of the
recent experimental results from the \babar\ and \belle\
collaborations. }

\section{Introduction}

The study of rare hadronic \B-meson decays with branching
fractions ${\cal O } (10^{-6})$ enable us to probe a wide range of
phenomena.  Such decays can be used to probe our understanding of
the CKM~\cite{CKM} description of quark mixing and \CP\ violation
(\CPV) in the Standard Model of Particle Physics (SM) by measuring
all three of the Unitarity Triangle angles $\alpha$, $\beta$ and $\gamma$, 
as well as charge asymmetries in direct \CPV\ $\Delta b = 1$ transitions. \CPV\
was first seen in the decay of neutral kaons~\cite{christenson}.
The first observation of \CPV\ in \B-meson decay was the
measurement of a non-zero value of $\sin2\beta$ by the
\B-factories~\cite{sin2beta}. The measurements of the Unitarity
Triangle angles and time-dependent \CP\ observables ($S$ and $C$)
in hadronic \B-meson decays are discussed elsewhere~\cite{uttalks}.
The phenomenon of direct \CPV\ was experimentally established in
the study of neutral kaon decay to two pion final states, through
the measurement of a non zero value of the ratio
$\epsilon^\prime/\epsilon=(1.67 \pm
0.26)\e{-3}$~\cite{epsilonprime,pdg}. In contrast to the kaon
system, where the direct \CPV\ parameter $\epsilon^\prime$ is
${\cal O}({\rm few}\e{-6})$, one can accommodate large direct
\CPV\ in \B-meson decay. For a \B-meson decays to a final state
$f$, the direct \CP\ asymmetry is defined as
\begin{equation}
\acp=(\overline{N} - N)/(\overline{N} + N),
\end{equation}
where $N$ ($\overline{N}$) is the number of B ($\overline{B}$)
decays to $f$ ($\overline{f}$). In order for a decay to be direct
\CP\ violating ($\acp\ne 0$), one requires that there are at least
two interfering amplitudes $A_i$ with both non-zero weak ($\phi$)
and strong phase ($\delta$) differences where
\begin{equation}
\acp\propto \sum_{i,j; i\ne j} A_iA_j \sin(\phi_i - \phi_j)
\sin(\delta_i - \delta_j).
\end{equation}
The values of the strong phases of amplitudes are {\em a priori}
unknown.  As a result, there is a limited amount of information
that one can extract from a measured \CP\ asymmetry in a single
decay channel.

There are still questions remaining on the subject of the
hierarchy and values of the branching fractions of a number of
rare hadronic \B-meson decays.  The \B-factories, the \babar\
experiment at SLAC~\cite{babar_nim} and the \belle\ experiment at
KEK~\cite{belle_nim}, continue to improve measurements of such
decays with the aim of providing theorists with more accurate data
to test calculations. There has recently been significant progress
on the experimental study of $\B \to hh$ ($h=\pi,K$) final states,
where the \babar\ experiment has accounted for the effects of
final state radiation in \B\-meson decays to $\pi^+\pi^-$,
$K^+\pi^-$, and $K^+K^-$~\cite{newpipibranchingfractions,fsrpipi}.
The recent observation of $\Bz\to a_1^+\pi^-$ might lead to
additional constraints on the value of the Unitarity Triangle
angle $\alpha$~\cite{gronauzupan2005}.

\B-meson decays to two vector particles ($V$) are particularly
interesting. In addition to the \CP\ violating observables, $S$,
$C$, and \acp, one has a number of angular correlation observables
to measure and compare with predictions. Any deviations from
theoretical expectations indicate either a deficiency in our
current understanding of these decays, or a tantalizing hint of
possible new physics effects. There has been considerable activity
in the study these $\B \to VV$ decays recently.

The remaining sections of these proceedings discuss experimental
techniques, results on searches for direct \CP\ violation,
branching fractions for the \B-meson decays to $h^+h^-$ and
$a_1^+\pi^-$, and studies of $\B\to VV$ decays. There is a summary
of results at the end of the article. Charge conjugation is
implied throughout these proceedings.

\section{Experimental techniques}

%
% continuum suppression
%
Signal candidates are identified using two kinematic variables
\mes\ and \deltae. The energy difference \deltae\ is defined as
the difference between the energy of the \B\ candidate and the
beam energy, $\sqrt{s}/2$, in the center of mass (CM) frame. The
beam-energy substituted mass, \mes, is defined as
 \begin{equation}
 \mes = \sqrt{(s/2 + {\mathbf {p}}_i\cdot {\mathbf {p}}_B)^2/E_i^2-
 {\mathbf {p}}_B^2},
 \end{equation}
where $\sqrt{s}$ is the total energy of both beams ($10.58 \gev$),
the \B\ momentum ${\mathbf {p}_B}$ and four-momentum of the
initial state $(E_i, {\mathbf {p}_i})$ are defined in the
laboratory frame.

Continuum $\epem \to \qqbar$ ($q = u,d,s,c$) events are the
dominant background to rare hadronic \B-decays.  To discriminate
signal from the continuum background one uses the fact that final
state particles in $B$ events tend to be spherically distributed,
whereas continuum events are more jet-like.  Event shape variables
are combined into a single variable with the purpose of
discriminating between signal and continuum background events (For
example, see Refs.~\cite{belle_pipi} and~\cite{babar_pipi}.).
%
% ML fit
%
All of the results discussed here are obtained using an extended
unbinned maximum-likelihood fit to the data.

\section{Results}
\subsection{Direct \CP\ violation searches}

\begin{figure}[!h]
\begin{center}
  \resizebox{8cm}{!}{\includegraphics{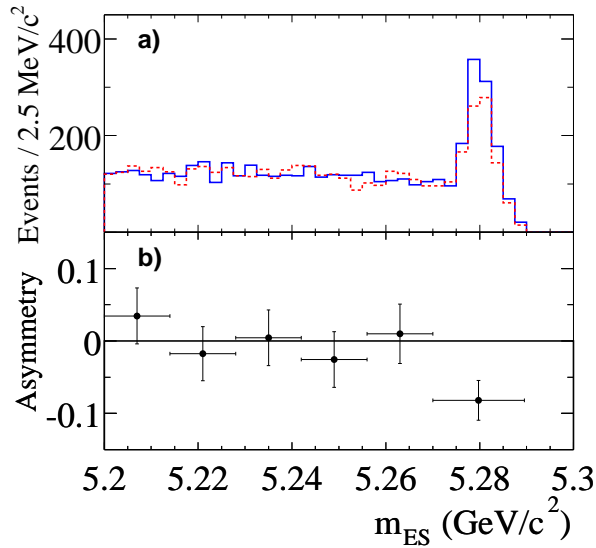}}
\end{center}
 \caption{Distributions of (a) \mes $\Bzb \to K^-\pi^+$, and $\Bz \to K^+\pi^-$
from the \babar\ data, and (b) \acp\ taken from
Ref.~\cite{babar_btokpi}. The solid histogram is for $K^+\pi^-$
and the dashed histogram is for $K^-\pi^+$.  The region below the
\B\ mass is dominated by continuum background. The \belle\ data
exhibit similar properties~\cite{belle_btokpi}.}
\label{fig:directcpv}
\end{figure}

\Bz-meson decays to the $K^+\pi^-$ and $K^-\pi^+$ final states are
self tagging in that the charge of the kaon in the final state
determines the flavor of the decaying \B-meson: $\Bzb \to
K^-\pi^+$, and $\Bz \to K^+\pi^-$. The \B-factories have measured
a significantly non-zero \acp\ in these decays.  The results
reported are
\begin{eqnarray}
 \acp\,_{K\pi} &=& -0.113 \pm 0.022 \stat \pm 0.08\syst, ({\mathrm {Belle}})\nonumber\\
 \acp\,_{K\pi} &=& -0.133 \pm 0.030 \stat\pm 0.09\syst, (\babar)\nonumber
\end{eqnarray}
using 386\e{6}, and 227\e{6} \bb\ pairs,
respectively~\cite{babar_btokpi,belle_btokpi}. On averaging the
results one obtains $\acp=-0.115 \pm 0.018$~\cite{HFAG}.  This
average includes results from the CDF and CLEO experiments, however the
\babar\ and Belle results dominate the average. This constitutes
an observation of direct \CPV\ in \B-meson decay, which is the
second type of \CPV\ observed in \B\ decays.
Figure~\ref{fig:directcpv} shows the \mes\ distributions of $\Bzb
\to K^-\pi^+$, and $\Bz \to K^+\pi^-$ from the \babar\ data.

There is considerable activity in searching for other possible
indications of direct \CPV\ at the \B-factories.  So far, there has
been no additional observations of this effect to date. The best
evidence obtained for direct \CPV\ in other \B\ decays is in the
channel $\Bz\to\pi^+\pi^-$.  More statistics are required to
establish \CPV\ in this decay. The Belle data give a $4.0\sigma$
evidence for direct \CPV~\cite{belle_pipi}.  The \babar\ data do
not yet show any evidence for direct \CPV~\cite{babar_pipi}.  The
measurements are
\begin{eqnarray}
\C_{\pi\pi} &=& -0.56 \pm 0.12 \stat \pm 0.06 \syst, ({\mathrm {Belle}})\nonumber \\
\C_{\pi\pi} &=& -0.09 \pm 0.15 \stat \pm 0.04\syst, (\babar)\nonumber
\end{eqnarray}
using 275\e{6}, and 232\e{6} \bb\ pairs, respectively.

Another promising decay channel that provides evidence for direct
\CPV\ is $B^+\to \rho^0 K^+$.  The Belle data are consistent with
a $3.9\sigma$ evidence for direct \CPV, whereas currently the
\babar\ data shows no indication of an asymmetry~\cite{btokpipi}.
The measured asymmetries are
\begin{eqnarray}
\acp\,_{\rho^0 K^+} &=& 0.30 \pm 0.11 \stat \pm 0.02 \syst^{+11}_{-4} ({\mathrm {model})}, ({\mathrm {Belle}})\nonumber \\
\acp\,_{\rho^0 K^+}  &=& 0.32\pm 0.13 \stat \pm 0.06 \syst^{+8}_{-5} {(\mathrm {model})}, (\babar)\nonumber
\end{eqnarray}
using 386\e{6}, and 226\e{6} \bb\ pairs, respectively.   Again,
more data are required in order to establish direct \CPV\ in this
mode.

\subsection{Branching fractions of $B\to hh$ decays}

The decay of \B\ mesons to $\pi\pi$, and $K\pi$ final states are
of general interest to the field of \B\ physics.  The $\Bz \to
\pi^+\pi^-$ branching fraction is an input to the $\pi\pi$ isospin
analysis, and measurement of $\alpha$~\cite{gronaulondon}, and
measurements of the decays to $K\pi$ are inputs to help elucidate
the so-called $K\pi$ puzzle~\cite{kpipuzzle}.

%The $\B \to h^+h^-$ branching fractions have been measured by the
%\B-factories on a fraction of current
%statistics~\cite{oldbtopipibranchingfractions}.
The \babar\ collaboration have recently updated these measurements
with 232\e{-6}\ \B pairs~\cite{newpipibranchingfractions}.  In this
analysis, \babar\ take into account the possible effects of final
state radiation (FSR)~\cite{fsrpipi}. Taking FSR into account has
ramifications on the extraction of the signal yield, as well as
the quoted signal efficiency that is used in calculating branching
fractions. The results obtained are:
\begin{eqnarray*}
 {\cal B}(\Bz \to K^+ \pi^-) &=& (19.2 \pm 0.6 \stat \pm 0.6 \syst)\e{-6},\\
 {\cal B}(\Bz \to \pi^+ \pi^-) &=& (5.5 \pm 0.4 \stat \pm 0.3 \syst)\e{-6},\\
 {\cal B}(\Bz \to K^+ K^-) &<& 0.4\e{-6} (90\% \, C.L.).
\end{eqnarray*}

The $K\pi$ ($\pi\pi$) branching fractions reported are 7\% (17\%)
higher than previous measurements from the
\B-factories~\cite{oldbtopipibranchingfractions}.  This highlights
the significance of further understanding and treatment of
FSR in rare hadronic \B-decays.

\subsection{Observation of the decay \Bztoaonepi}

Some time ago it was suggested that one could measure the
Unitarity Triangle angle $\alpha$ using the non-\CP\ eigenstate
decay \Bztoaonepi~\cite{aleksanaonepi}.  \babar\ has recently
observed the decay \Bztoaonepi, and Belle have confirmed this
observation~\cite{btoaonepi}. The measured branching fraction for
this decay is
\begin{eqnarray*}
 \Br(\Bztoaonepi)&=&(33.2 \pm 3.8\stat \pm 3.0\syst)\e{-6}, (\babar) \nonumber \\
 \Br(\Bztoaonepi)&=&(48.6 \pm 4.1\stat \pm 3.9\syst)\e{-6}, {(\mathrm {Belle})}
\end{eqnarray*}
using 218\e{6}, and 275\e{6} \bb\ pairs, respectively.  One can
expect the \B-factories to investigate the prospects of such a
time-dependent \CP\ analysis in the coming years. An isospin
analysis of $\B\to\aone\pi$ decays would be experimentally
challenging. The use of \su{3}\ to measure $\alpha$ with $\B \to
a_1 \pi$ decays has recently been proposed~\cite{gronauzupan2005}
as an alternative.  This approach requires experimental knowledge
of the decays $\B\to \aone K$, $\B\to K_1(1270)\pi$ and $\B\to
K_1(1400)\pi$. \babar\ recently performed a search for the related
decay $\Bz\to\aone^\pm\rho^\mp$~\cite{btoaonerho} using 110\e{6}
\bb\ pairs.  The result of this search was an upper limit of
$<61\e{-6}$ (90\%C.L.).

\subsection{Dynamics of \B-meson decays to two vector particles}

\begin{figure}[!h]
\begin{center}
 \resizebox{8cm}{!}{\includegraphics{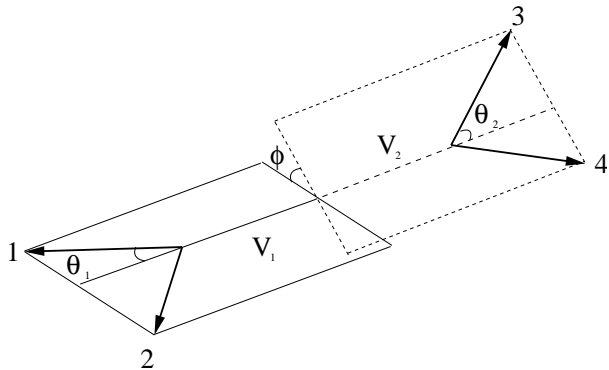}}
\end{center}
 \caption{The decay of a \B-meson to via two vector mesons, $V_1$ and $V_2$, to a four particle final state.}
\label{fig:btovv}
\end{figure}

\begin{figure}[!h]
\begin{center}
  \resizebox{8cm}{!}{\includegraphics{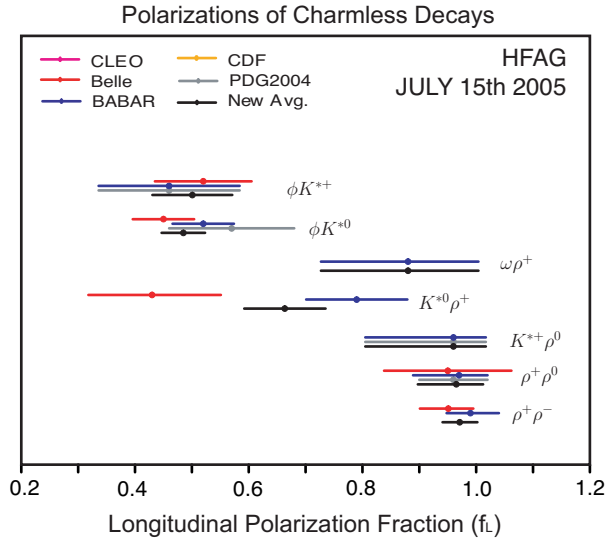}}
\end{center}
 \caption{The measured values of \ptrue\ in $B \to VV$ decays~\cite{HFAG}.}
\label{fig:btovvresults}
\end{figure}

The \B-factories have performed studies of the angular
correlations of several types of $B\to VV$ decay: $B\to \rho\rho$,
$\B \to K^\star\rho$, $\B \to K^\star\omega$, and $\B \to
K^\star\phi$~\cite{btovv,btophikstar}.
 Figure~\ref{fig:btovv} shows a schematic of the topology of a
$B\to VV$ decay, where each vector particle decays to a two body
final state.
where $\theta_{i}$ (i=1,2) is the helicity angle of the vector
particle defined as the angle between the daughter momentum in the
vector particle rest frame and the flight direction of the \Bz\ in
this frame.  The angle $\phi$ is the angle between the decay
planes of the vector mesons. Most of the decays studied have
limited statistics and focus on extracting the fraction of
longitudinally polarized events ($\ptrue$) from the data after
integrating over $\phi$. The angular distribution used for these
decays is
\begin{eqnarray}
&&\frac{d^2\Gamma}{\Gamma d\cos\theta_1 d\cos\theta_2}=
\frac{9}{4}\left(f_L \cos^2\theta_1 \cos^2\theta_2 +
\frac{1}{4}(1-f_L) \sin^2\theta_1 \sin^2\theta_2 \right),
\end{eqnarray}
A full angular analysis is performed for \B-meson decays to $\phi
K^*$~\cite{btophikstar,Dunietz}. Early calculations of \ptrue\
predicted the longitudinal polarization would dominate in $\B\to
VV$ decays~\cite{Suzuki}. Figure~\ref{fig:btovvresults} shows the
measured values of \ptrue\ obtained from experiment.
There is a pattern to the underlying physics processes which needs
to be understood in the measured values of \ptrue. The tree
dominated decays of \B\-mesons to $\rho\rho$, $K^\star\omega$ and
$K^\star\rho$ are consistent with the naive expectation that:
$\ptrue \sim {\cal O}(1 - m^2_V/m^2_b)$. However the loop
dominated modes ($K^\star \phi$ and $K^{\star +}\rho^-$) have
lower values of \ptrue\ than expected. It is possible that
improvements in calculations of \ptrue\ for the loop dominated
modes could resolve this issue.  Possible new physics
contributions modifying these observables have also been
discussed~\cite{btovvtalks}. There remains work to be done in this
area, both on the experimental and theoretical side.  The
\B-factories need to perform as many different measurements in as
many different $\B \to VV$ decay modes as possible.  One can 
also measure \CP-violating, \T-odd triple product asymmetries in 
$B\to VV$ decays~\cite{londonbtovv}.

\section{Summary}

The \B-factories have established direct \CP\ violation in $\Bz
\to K^\pm \pi^\mp$ decays in recent years.  Tantalizing hints of
direct \CPV\ are starting to emerge in other modes. As the
\B-factories continue to accumulate data we should see direct
\CPV\ being established in more decay modes.

In summary, the study of rare hadronic \B-meson decays has
provided a rich harvest of information since the \B-factories
started taking data.  There is still a wide range of physics that
we can learn from using these decays.  The recent observation of
\Bztoaonepi\ raises the possibility that the \B-factories might be
able to add another mode to the list of those providing
measurements of the Unitarity Triangle angle $\alpha$.
Experimental and theoretical work is required to understand the
pattern of measured \ptrue\ in $B\to VV$ decays.

\section{Acknowledgments}

This work is supported in part by the UK's Particle Physics and
Astronomy Research Council, and the U.S. Department of Energy
under contract number DE-AC02-76SF00515.

%%%%%%%%%%%%%%%%%%%%%%%%%%%%%%%%%%%%%
% Label to flag the last page of your contribution
% Replace Perret by your name starting with a capital letter
%
\label{AJBevanEnd}

\end{document}